# An alternative interpretation for the Holm "alpha model"

David C. Montgomery[*] and Annick Pouquet[†]
*Advanced Study Program,
National Center for Atmospheric Research,
P.O. Box 3000, Boulder, Colorado 80307*

By re-interpreting a recent successful closure procedure in terms of local spatial averaging and the neglect of fluctuations about that average, it is shown how the results of that closure scheme (the "alpha model") may be more simply derived.

Great effort over the last half-century has been expended on devising "closure" schemes for the Navier-Stokes equations, so that incompressible fluid turbulence computations can be carried out at high Reynolds numbers (as encountered, e.g., in geophysical and astrophysical turbulent flows) with a reduced number of fluid variables. One of the most promising closures of recent years has been the so-called "alpha model" of Holm and collaborators (see, e.g., Refs. [1] [2], each of which contains an extensive bibliography). The scheme has demonstrated impressive results when tested against laboratory data [3] and the output of direct Navier-Stokes numerical solutions [4].

The model has been developed in the framework of statistical-mechanical averaging, either over ensembles (of Lagrangian trajectories), over time or over phases. It has also been linked with abstract and somewhat forbidding mathematics. In particular, a strong connection has been made explicit with the GLM (Generalized Lagrangian Mean) of Andrew and McIntyre (1978). Further manipulations – namely, (i) considering small–scale equations, (ii) introducing a Taylor hypothesis (of sweeping the small scales by the large scale flow) and (iii) isotropy of the resulting model – lead to a particularly simple equation that bears resemblance to the original Navier–Stokes equations and introduces an arbitrary parameter $\alpha$, a length scale.

The purpose of this note is to show how an alternative interpretation of the model leads to the same mathematical results by using only elementary and easily-accessible mathematics. Thus, there are no really new equations presented here, only a re-derivation of known ones by simpler means.

We start with the Navier-Stokes equation in the form,

$$\partial_t \mathbf{v} + \mathbf{v} \cdot \nabla \mathbf{v} = -\nabla p + \nu \nabla^2 \mathbf{v} , \qquad (1)$$

where $\mathbf{v} = \mathbf{v}(\mathbf{x}, t)$ is the (solenoidal) fluid velocity, $p = p(\mathbf{x}, t)$ is the pressure normalized to unit mass density, and the kinematic viscosity is $\nu$. We will drop the viscous terms for simplicity and work with the ideal Euler equations, though they may be re-inserted at any time. We also work in the infinite domain, for simplicity, and further use the vorticity representation, by taking the curl of Eq. (1) and using the fact that $\mathbf{v}$ is divergenceless,

$$\partial_t \boldsymbol{\omega} + \mathbf{v} \cdot \nabla \boldsymbol{\omega} = \boldsymbol{\omega} \cdot \nabla \mathbf{v} , \qquad (2)$$

where the vorticity vector is $\boldsymbol{\omega} = \nabla \times \mathbf{v}$.

For any particular turbulent solution, we define a locally smoothed velocity distribution $\mathbf{u}(\mathbf{x}, t)$ by the integral relation,

$$\mathbf{u}(\mathbf{x}, t) = \int G_\alpha(\mathbf{x} - \mathbf{x}') \mathbf{v}(\mathbf{x}', t) \, d^3\mathbf{x}' \qquad (3)$$

where $G_\alpha$ is given by

$$G_\alpha(\mathbf{r}) = \frac{exp[-|\mathbf{r}|/\alpha]}{4\pi \alpha^2 |\mathbf{r}|} = \int \frac{exp[i\mathbf{k} \cdot \mathbf{r}]}{1 + k^2 \alpha^2} \frac{d^3\mathbf{k}}{(2\pi)^3} . \qquad (4)$$

The characteristic length scale over which $\mathbf{v}$ is smoothed is left arbitrary, and is called $\alpha$.

We define the local fluctuation about $\mathbf{u}$ by $\delta \mathbf{v} = \mathbf{v} - \mathbf{u}$, and rewrite Eq. (2) in terms of $\mathbf{u}$ and $\delta \mathbf{v}$ as

$$\partial_t \boldsymbol{\omega} + (\mathbf{u} + \delta \mathbf{v}) \cdot \nabla \boldsymbol{\omega} = \boldsymbol{\omega} \cdot \nabla (\mathbf{u} + \delta \mathbf{v}) , \qquad (5)$$

which is still exact.

If we now assume $\delta \mathbf{v}$ to be small compared to $\mathbf{u}$ and neglect it in Eq. (5), the immediate result is:

$$\partial_t \boldsymbol{\omega} + \mathbf{u} \cdot \nabla \boldsymbol{\omega} = \boldsymbol{\omega} \cdot \nabla \mathbf{u} . \qquad (6)$$

This neglect of $\delta \mathbf{v}$ relative to $\mathbf{u}$ is the only approximation necessary anywhere in the development.

Eq. (6), together with Eqs. (3) and (4), is just the alpha model. If we remove a curl from Eq. (6), use a well-known vector identity, and re-arrange the terms, the result is:

$$\partial_t \mathbf{v} + \mathbf{u} \cdot \nabla \mathbf{v} + v_j \nabla u^j + \nabla P = 0 , \qquad (7)$$

where $P$ is immediately recognizable as a pressure-like scalar which must be determined from the Poisson equation for it, obtained by taking the divergence of Eq. (7)

---

[*]Also at Department of Physics and Astronomy, Dartmouth College, Hanover, NH 03755; Electronic address: David.Montgomery@Dartmouth.edu
[†]Electronic address: pouquet@ucar.edu

in the usual way. Eq. (7) is one among several equivalent forms in which the alpha model can be written [1, 2], with the differential connection between **u** and **v** given by

$$\mathbf{v} = (1 - \alpha^2 \nabla^2)\mathbf{u} \ . \tag{8}$$

It should be noted that a different kernel or Green's function than the one written in Eq. (4) could have been used with the same formal result, but that the choice of Eq. (4) or, equivalently, Eq. (8), provides a particularly desirable connection between **u** and **v** in Fourier transform space, where many computations are actually carried out. The dynamics can, of course, be rewritten in a variety of ways, including entirely in terms of the smoothed **u** field without further approximation or assumption [1, 2]. Viewed in Fourier space, the smoothed velocity field **u** is obtained through partially suppressing the small scales present in a turbulent flow, and indeed allows for control of these scales. Viewed in configuration space as given by Eq. (4), it shows that such suppression stems as well from a local approximation, eliminating far–away (with respect to $\alpha$) spatial contributions.

Note from Eq. (4) that the effect of the filter is not a complete suppression of the small scales, but permits some "leakage" to them. (As alpha approaches zero, G approaches a delta-function filter.) A somewhat steeper spectrum than the classical Kolmogorov one, beyond the wave number corresponding to the reciprocal of $\alpha$, will result as a consequence of the partial suppression of the small scales. In large eddy simulations (LES), filters with no leakage have also been used, as have those with exponential decays; but so have spatial "box" filters, which have less rapid Fourier space decays. See Refs. [6] [7] for more comprehensive descriptions of properties of LES filters.

It remains an "experimental" question as to how well the reduced description of the alpha model reproduces the desired properties of true Navier-Stokes solutions. Considerable impressive evidence of its ability to do so has accumulated already, with more likely to come, and we do not undertake the task of reviewing that evidence here.

What has been done, rather, is to show how the alpha model's mathematical results follow, interpreted for one fluid rather than for an ensemble, as a consequence of two simple steps: a local spatial averaging of the velocity field followed by the neglect of the fluctuation about that average in the vorticity-representation dynamics. An identical program can be readily carried through for one-dimensional Burgers–like equations, as well as for magnetohydrodynamics (MHD), including the Hall and ambipolar drift terms in a generalized Ohm's law, and for two-dimensional turbulence.

Two features of the above development may be worth further mention. First, notice that the velocity field **v** , but not the vorticity field associated with **v** , has been subjected to the smoothing operation. This certainly does not decrease the accuracy, but introduces an asymmetry in the way the two fields are treated that might usefully be elucidated. Secondly, the possibility arises that the alpha model might under some circumstances be even better than our own approximation suggests. For the terms in the velocity fluctuation that are discarded from the exact vorticity evolution equation may in some circumstances be rapidly varying in time, and thus may influence the convected vorticity less than a formal order-of-magnitude estimate would indicate.

We are grateful to Dr. Darryl Holm for a thorough presentation to us of the alpha model and some of its intriguing consequences.